\documentclass[10pt,conference]{IEEEtran}
 
\usepackage{float}
\usepackage{epsfig}
\usepackage{amsmath}

\ifCLASSINFOpdf
\else
\fi
\begin{document}
%
\title{Continuous Molecular Communication in one dimensional situation}

\author{\IEEEauthorblockN{Pengfei Lu}
\IEEEauthorblockA{School of Computer Science\\
Shaanxi Normal University\\
xi'an, China, 710119\\
Email: jacinto@snnu.edu.cn}
\and
\IEEEauthorblockN{Zhenqiang Wu}
\IEEEauthorblockA{School of Computer Science\\
Shaanxi Normal University\\
xi'an, China, 710119\\
Email: zqiangwu@snnu.edu.cn}
}

%


\maketitle

\begin{abstract}
Molecular Communication as the most potential methods to solve the communication in nano scale, for it's derived from nature, and it becomes more and more prevalent. Though molecular communication happens in three dimensional situation, there are also some situation that are in the one dimensional situation, especially when considering the transmitters and the receivers are in extremely short distance or in long slim pipe. In this paper, we introduce the one dimensional situation, and studied how the continuous information molecules transmitted in this situation, also introduced how to encode and decode the information molecules, and based on the molecular communication model, we studied some metrics of it, such as the distance between transmitter and receiver, the emitting frequency of transmitter. Through the research we know that the distance and frequency are important metrics to the successful communication, which can direct us how to place the nano transmitters and receivers in the future nano network environment.
\end{abstract}


%
\IEEEpeerreviewmaketitle

\section{Introduction}
With the quickly development of nano technology, the scale of electrical devices becoming smaller and smaller, and the ability of single nano machine is very limited due to its size, therefore, sharing information among nano machines is good for the use of single nano machine through the connection of them, which organise the nano network. Achieving the communication of nano machines has two main ways which are nano electrical communication through Carbon nanotubes and molecular communication through information molecules. Among them, molecular communication(MC) is the most potential way to achieve, which relies on the bio-inspired method and fits for the biological environment.\\
At the open space of biological environment, the three dimensional space is fit for molecular communication\cite{2013-Nakano-p-} to model channel. In Mahfuz et al's articles, they elaborated on the spatiotemporal distribution of information messages under the binary concentration-encoded modulate methods. Mahfuz's researches have a great impact on the study of molecular communication, which gives us a new modulate methods. For the binary concentration-encoded modulate methods, it's much fit for the biological environment, and without considering the inner structures of the molecules. 
Also in Akan's paper, he studied the ligand and receptor process of molecular communication, in order to study the details of how molecules bind with the receptors, which is based on the binary concentration-encoded modulate methods.\\
All of their  works help us understand the molecular communication clearly. As we know in molecular communication different modulation methods or different transportation environment can make different channel. Most studies are based on the three dimensional environmental, and fews are concentrate on the one dimensional environment. Comparing with the three dimensional , one dimensional can make sure the receiver receive all the information molecules for a long time, and the one dimensional situation is simple enough, therefore, it is worth for us to study it and it has a wealth of applications. For instance, in Nariman's molecular communication test-bed, which is studied based on the one dimensional environment. Also in extremely tiny pipe, the transmitter and receiver nano machines on sides of the pipe can regard as in the same line. Additionally, when transmitter and receiver nano machines are in extremely small distance, we can consider they are in one dimensional environment.\\
In this paper, we mainly considered in the tiny slim pipe, how the continuous information molecules are transmitted. And how the information transmission is affected by the transmission frequency and the distance of transmitter and receiver. The study can help us know how the frequency affect the communication between nano machines. Also can help us to place the nano machines in the future.
\\
The paper is organised as follows: in Section \ref{sec_mc_model}, we produced a molecular comunication model; in Section \ref{sec_sample_threshold} we introduced the sample method and the threshold of the receiver; in Section \ref{sec_analysis}, we analyzed some parameters which affect the model; in Section \ref{sec_conclusion}, we make a conclusion of our work.
\section{Molecular Communication via Diffusion}\label{sec_mc_model}
\begin{figure}[H]
		\centering
		\noindent\makebox[0.5\textwidth][c]{
		\includegraphics[width=0.4\textwidth,height=0.23\textwidth]{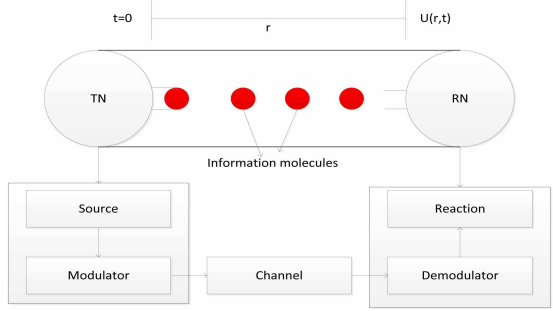}
		}
		\caption{The molecular communication system}
		\label{fig_mc_model}
\end{figure}

\begin{figure}[H]
	\centering
	\noindent\makebox[0.5\textwidth][c]{
	\includegraphics[width=0.4\paperwidth,height=0.25\paperwidth]{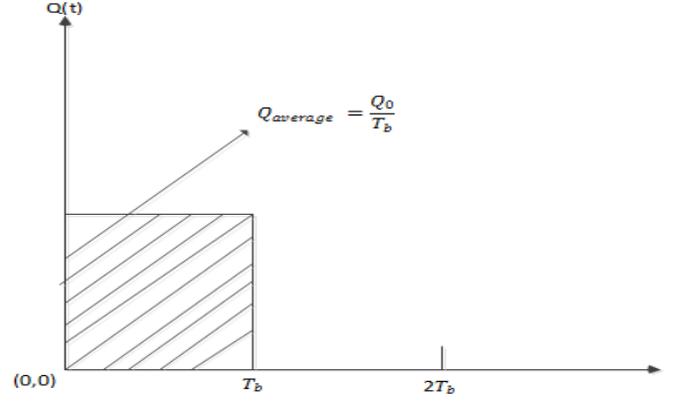}
	}
	\caption{ON-OFF modulation with constant amplitude pulse transmission\cite{2010-Mahfuz-p289-300}.}
	\label{fig_on_off_modulation}
\end{figure}

We model a communication system composed of a pair of devices which are connected by an extremely slim pipe, each called a nanonetworking-enabled node(NeN,ie., nano node or nano robot)\cite{2014-Yilmaz-p929-932}. The communication system is described as Fig.\ref{fig_mc_model}. In molecular communication via diffusion(MCvD), the NeNs communicate with each other through the propagation of certain molecules via diffusion\cite{2010-Pierobon-p602-611}. The TN transmits the encoded information molecules in the propagation medium, the communication phase being known as the sending phase. The transmitted molecules propageated by the channel, the bind with the RN ,which is known as the ligand-receptor binding\cite{2010-Mahfuz-p289-300}. Since the TN and RN nano machine are connected by the extremely slim pipe, surely, we can consider the TN and RN in the one dimensional environment, then we can study its spatiotemporal distribution of signal strength in one dimensional environment.
The emission of information can be an instantaneous or a continuous emission. As the instantaneous release of information molecules, in a single puff, would be ideal design for communications requiring rapid fade-out. On the other hand, the source emitting message molecules continuously at a constant rate, to some extent might be useful for status telemetry, navigational beacons or periodic sampling monitors\cite{2009-Lacasa-p-}. In this paper we regard the emission is continuous.\\
The spatiotemporal distribution of molecules transmitted by the TN that will be available at RN is calculated by the well-known Roberts equation explained in \cite{1963-Bossert-p443-469}. We assume that a point source type TN continuously emitting molecules at a rate $Q(t)$ molecules per second, x is taken as the distance down the pipe from the source, $A$ is the cross-sectional area of the long and slim pipe,and the concentration of molecules $U(x,t)$ at distance $x$ and time instant $t$, then in one dimensional situation, $U(x,t)$ is described as follows Eq.\ref{equation_spatiotemporal_1D}, 
\begin{equation}\label{equation_spatiotemporal_1D}
U(x,t)=\int_0^t{\frac{Q(t)}{A\sqrt{4\pi{D}(t-t^{\star})}}e^{\frac{-x^2}{4D(t-t^{\star})}}}dt^{\star} \quad t>0.
\end{equation}
where $t^{\star}$ as the dummy variable of integration, D is the diffusion constant in $cm^2/sec$.\\
We use the ON-OFF modulation method to transmit information. By transmit 1 bit, the TN will emitting a quantity of molecules $Q(Q>0)$, and transmit 0 bit, the TN will emitting nothing to the channel. We utilize $f_b$ stands fot the emitting frequency of TN, and $T_b=1/f_b$ is the emitting period. In the emitting period, we can emitting molecules continuously, or transmitting nothing as described in Eq.\ref{equation_on_off} or Fig.\ref{fig_on_off_modulation}.
\begin{equation}\label{equation_on_off}
Q(t)=\left\{
\begin{aligned}
Q_{average}; \quad \quad \text{for bit '1'}\\
0; \quad \quad \text{for bit '0'}
\end{aligned}
\right.
\end{equation}
\section{Sample method and concentration threshold}\label{sec_sample_threshold}
\begin{figure}[H]
	\centering
	\noindent\makebox[0.5\textwidth][c]{
		\includegraphics[width=0.4\paperwidth,height=0.25\paperwidth]{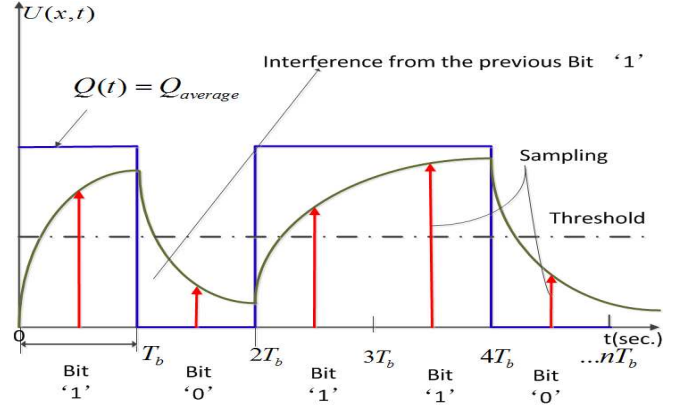}
	}
	\caption{spray in single duration\cite{2011-Mahfuz-p446-449}}
	\label{fig_single_duration_spray}
\end{figure}

In this paper, we regard $Q(t)$ as a constant value $Q_{average}$ which is the average molecules in the emitting period. As aforementioned we transmit "1" when emitting a quantity of molecules, and transmit "0" when emitting nothing. In this situation, the distance between TN and RN is x, as Fig.\ref{fig_single_duration_spray} describes, we try to transmit bit sequences "10110".
When molecules were transmitted to the receiver(RN), the RN needs to decide the information is "1" or "0. Therefore, we need to set a threshold for the receiver, by comparing the available information molecules(AIM) to the threshold, we can know the information is "1" if the AIM is great or equal than the threshold, and in vice versa, the information is "0". In the RN side, it's no need to sample the value of AIM for each time instant, according to the Nyquist–Shannon sampling theorem, we can get the sampling value at time instant of $\frac{(2n-1)T_b}{2}$ is $Z_{SD}$, where $n$ is the length of the bit information, we only need to compare the $Z_{SD}$ to the threshold.
In Fig.\ref{fig_single_duration_spray}, the red arrow is the sampling instant, and the interference is the noise information molecules from the previous molecules informations, in this paper, for a simple description, we only need to consider one bit information of the previous information molecules.
\section{Analysis}\label{sec_analysis}
In this section, we study how the distance and emitting frequency affect the transmission under the environment of air medium. And at last, we do the simulation to testify the difference of information transmission in one dimensional(1-D) situation and three dimensional(3-D) situation. In this section, we only transmit two bit information "10", for it can help us analysis the AIM in the first bit "1", and interference in the second bit "0".
\subsection{How distance affect the concentration}
\noindent We assume that the $Q_{average}$ is 10000 molecules/sec, $T_b$ is 30$sec$, the diffusion coefficient in air medium is $D$=0.43$cm^2/s$, we utilize the model in Eq.\ref{equation_spatiotemporal_1D} to do the simulation at different distances, such as $x$ are 0.05$cm$,1$cm$,10$cm$ respectively. 
\begin{figure}[H]
	\centering
	\noindent\makebox[0.5\textwidth][c]{
	\includegraphics[width=0.4\paperwidth,height=0.25\paperwidth]{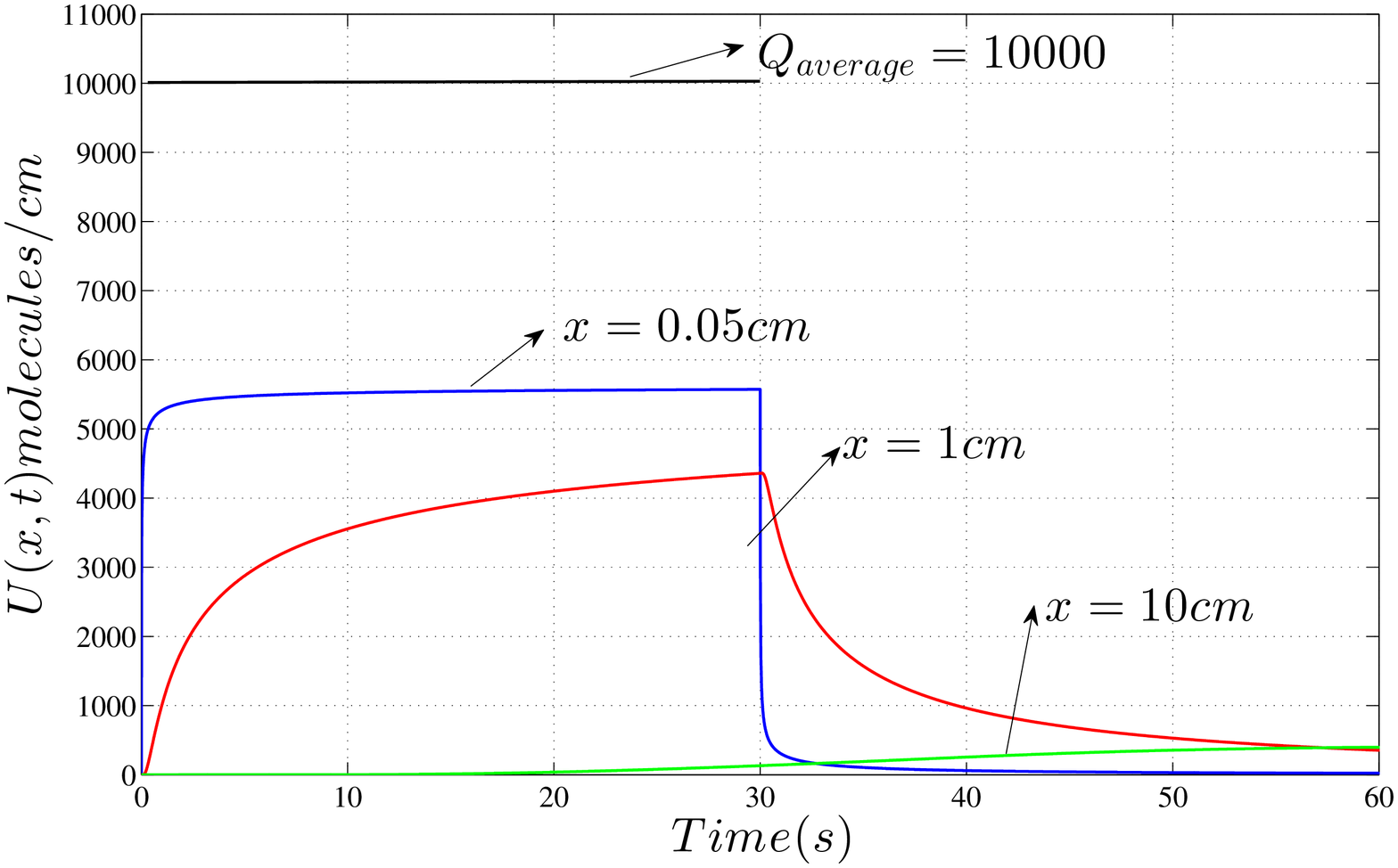}
	}
	\caption{The available concentration in different distance $x$.}
	\label{fig_distance_varies}
\end{figure}
Through the Fig.\ref{fig_distance_varies}, we find that with the increase of distance between TN an RN, the AIM in the RN side is become smaller and smaller, but the interference is becoming larger and larger. Therefore, we need to set the distance at short distance, in order to transmit enough information molecules, and improve the probability of successful information transmission.
\subsection{How frequency affect the concentration}
Assuming that the $Q_{average}$ is 10000 molecules/sec, the distance between TN and RN is $d=1cm$, and the diffusion coefficient in air medium is $D=0.43cm^2/s$. Using the model in Eq.\ref{equation_spatiotemporal_1D}, we do the simulation at different emitting frequencies, such as  $\frac{1}{10}Hz$,$\frac{1}{20}Hz$ and $\frac{1}{30}Hz$.
\begin{figure}[H]
	\centering
	\noindent\makebox[0.5\textwidth][c]{
	\includegraphics[width=0.4\paperwidth,height=0.25\paperwidth]{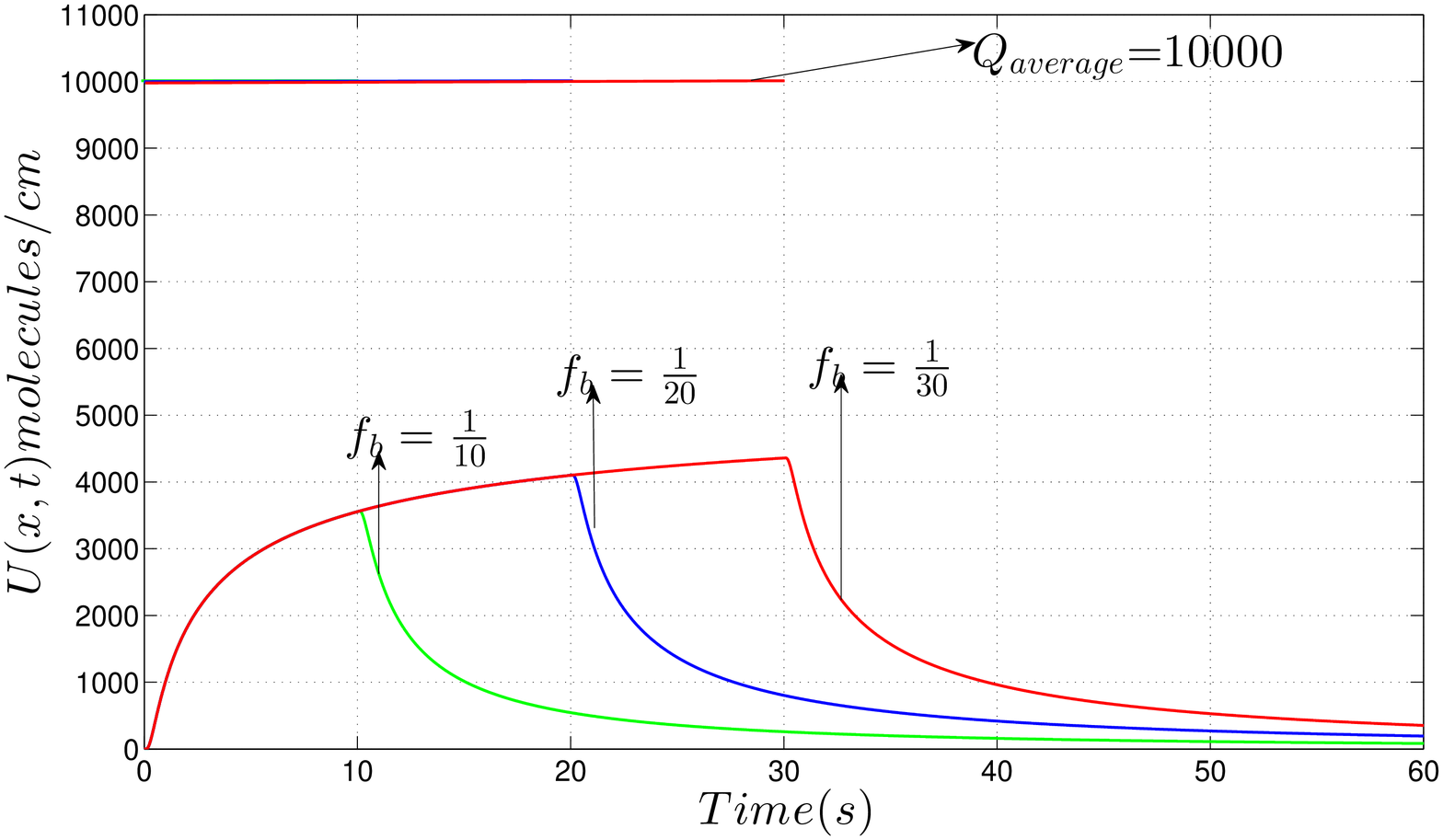}
	}
	\caption{The available concentration in different frequency $f$}
	\label{fig_frequent_varies}
\end{figure}
Through the Fig.\ref{fig_distance_varies}, we find that with the increase of $f_b$, the AIM in the RN side is become smaller and smaller, also the interference is becoming smaller and smaller. Since the AIM and interference increase or decrease simultaneously, we need to make a compromise for the change of frequency, in order to transmit enough information molecules, and improve the probability of successful information transmission.
\subsection{The comparison available of molecule concentration in 1-D and 3-D}
\begin{figure}[H]
	\centering
	\noindent\makebox[0.5\textwidth][c]{
	\includegraphics[width=0.4\paperwidth,height=0.25\paperwidth]{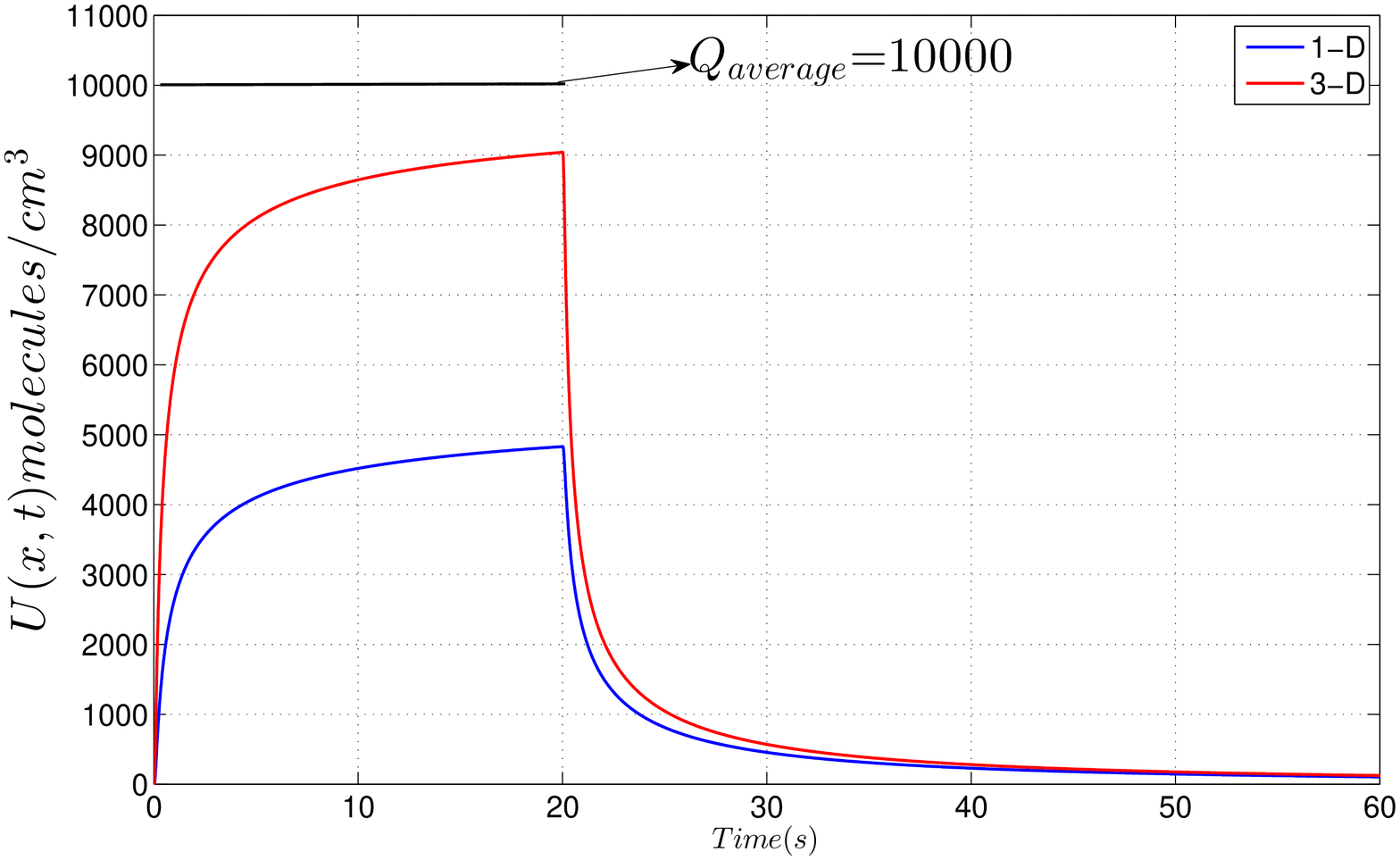}
	}
	\caption{The comparison of molecules transmission in 1-D and 3-D medium}
	\label{fig_compare_1D_3D}
\end{figure}

We need to do some researches to study the differences between one dimensional situation and three dimensional situation. We use the same basic values of the parameters described before, except for the distance $x=0.5$, the frequency $f=\frac{1}{20}$. The three dimensional model of molecules transmission is adapted from \cite{2015-Mahfuz-p67-83}. The simulation result shows in Fig.\ref{fig_compare_1D_3D}.
From Fig.\ref{fig_compare_1D_3D}, we know that the 3-D line higher than 1-D, and the interference is almost the same. In one dimensional situation can make sure the receiver the information molecules completely in certain time, however, the available information in three dimensional situation is higher. We think the reason is that in three dimensional situation, there are a large quantities of molecules in certain time slot, but to the one dimensional situation the information molecules needs to transmit sequencely which result the through out is smaller.
\section{Conclusion}\label{sec_conclusion}
In this paper, we studied continuous molecules were transmitted in one dimensional molecular communication. We use binary concentration-encoded method to encode information molecules into bit "1" or "0", also we introduce sample theorem and threshold for the receiver to decode information molecules. To study the molecular communication deeper, we studied that how the metrics of distance and frequency affect the molecules concentration, and we learned that the longer the distance the lower of the available information molecules in RN, but higher in interferences, and the frequency has two sides on the available molecular communication, either positive or negative, therefore, we should make a compromise between the available information molecules and the interferences from the former information bits. Finally, we made a comparison between one dimensional situation and three dimensional situation, and we find that in three dimensional situation, the RN side has a higher available information molecules, but cannot make sure the RN receives the molecular information for one hundred percent.\\ 
Still there are many areas need to study, for the future research work, we can promote our research from three aspects. For the first,we can study how the variation of different emit rates affect the molecules concentration. For the second , we should find a method to make a cut-off between the frequency and the interference, in order to make the communication more accurate and high efficiency. For the third, we should consider to combine the advantages of the three dimensional and one dimensional situation to improve the available information molecules.


\section*{Acknowledgment}
This work was supported by the National Natural Science Foundation of China (61173190) and the Fundamental Research Funds for the Central Universities(GK201501008).



%
%
%



\end{document}